\documentclass[prd,twocolumn,showpacs,preprintnumbers,superscriptaddress,floatfix,nofootinbib]{revtex4-1}

\usepackage{subfigure}
\usepackage{amsfonts}
\usepackage{graphicx,booktabs}
\usepackage{amssymb}
\usepackage{amsmath,txfonts,ulem}
\usepackage{graphicx}
\usepackage{epstopdf}
\usepackage{epsfig}
\usepackage{dcolumn}
\usepackage{xcolor}
\usepackage{soul}
\usepackage{bm}
\usepackage{multirow}
\usepackage{ulem}
\usepackage[colorlinks,
citecolor=blue,
anchorcolor=green,
menucolor=orange,
linkcolor=red,
filecolor=red,
runcolor=pink,
urlcolor=blue,
frenchlinks=red]{hyperref}
\usepackage[figuresright]{rotating}
\usepackage{afterpage}
\usepackage{subfigure}
\usepackage{amsmath}
\usepackage{cases}
\usepackage{ulem}
\usepackage{mathrsfs}
\usepackage{booktabs}
\usepackage{amsmath}

\allowdisplaybreaks[4]
\begin{document}
	
	\title{Analysis of $J/\psi$ and $\psi(2S)$ Charmonium Production in Ultraperipheral Lead-Lead and Proton-Lead Collisions at LHC Energies}

	\author{Zhe Wang}
    \email{explorewang@yeah.net}
	\affiliation{Department of physics, Lanzhou University of Technology, Lanzhou 730050, China}

    \author{Jiyuan Zhang}    \email{jiyuanzhang24@m.fudan.edu.cn}
	\affiliation{Institute
       of Modern Physics, Fudan University, Shanghai 200443,  China}

    \author{Xiao-Yun Wang}
    \email{xywang@lut.edu.cn (Corresponding author)} 
    \affiliation{Department of physics, Lanzhou University of Technology, Lanzhou 730050, China}
    \affiliation{Lanzhou Center for Theoretical Physics, Key Laboratory of Theoretical Physics of Gansu Province, and Key Laboratory of Quantum Theory and Applications of MoE, Lanzhou University, Lanzhou, Gansu 730000, China}

	\begin{abstract}
The two gluon exchange model serves as a key framework for describing the photoproduction of heavy vector mesons, and the photoproduction cross sections derived from it provide essential input for studies of Ultra-Peripheral Collisions (UPCs). Building on the model’s successful description of $J/\psi$ and $\psi(2S)$ photoproduction in previous work, we use the STARlight program to systematically investigate charmonium UPCs in Pb–Pb and $p$–Pb collisions at LHC energies. To reduce discrepancies between theoretical predictions and experimental data for rapidity and transverse momentum distributions in Pb–Pb collisions, a phenomenological suppression factor is introduced to correct the theoretical results. We find that, in the TeV energy range, the corrected predictions agree well with experimental data within uncertainties and successfully reproduce the characteristic double-peak structure in rapidity distributions. In contrast, no significant suppression is observed in $p$–Pb UPCs, which reflects the asymmetric photon fluxes and the dominant contribution from the photon–proton interaction branch. The transverse momentum distributions from STARlight simulations also match the diffractive pattern of coherent production seen experimentally, although the overall yield remains slightly overpredicted. This work further validates the applicability of photoproduction cross sections from the two gluon exchange model for charmonium UPC studies, and offers valuable phenomenological guidance for future experimental design and data analysis in UPC measurements of charmonium production.
\end{abstract}
	
	
	\maketitle
	
	\section{Introduction}
	\label{sec1}

 	In 1974, particle physics achieved a major breakthrough: Samuel C. C. Ting's experimental group, using a proton accelerator to bombard a beryllium target, observed a new particle for the first time and named it the $J$ particle \cite{refn1}. Similarly, the Richter experimental group detected a significant resonance structure in the center-of-mass energy region of 3.1 GeV/$c^{2}$, naming it the $\psi$ particle \cite{refn2}. Two independently observed particles possess identical $c\bar{c}$ quark configurations, corresponding to the same particle, namely the $J/\psi$ particle. As crucial experimental evidence for the existence of the charm quark, the discovery of the $J/\psi$ particle significantly advanced the theoretical framework of the Standard Model and propelled the development of particle physics. As a direct follow-up to the $J/\psi$ discovery, the Mark II collaboration at SLAC achieved a breakthrough in 1974 by identifying the excited state $\psi(2S)$. This finding was formally documented in their 1975 publication \cite{refn3}, which established $\psi(2S)$ as the first radial excitation of charmonium system. The experimental identification of higher-energy charmonium states provided critical insights into the interplay between perturbative and non-perturbative QCD regimes, the formation mechanisms of quark-antiquark bound states, and the validity limits of potential models in heavy quark systems \cite{Brodsky:1994kf,refn4,refn5}.

    In addition to studying the intrinsic properties of charmonium particles themselves \cite{ParticleDataGroup:2024cfk}, researchers also utilize their decay or production processes to investigate important physical problems in hadron spectroscopy, nucleon structure, and other areas. For example, BESIII has accumulated the world's largest data samples of $J/\psi$ and $\psi(2S)$  \cite{BESIII:2021cxx, BESIII:2024lks}, which provide a unique hunting ground for glueballs, hybrids, and exotic multiquark states through radiative and hadronic decays \cite{Chen:2022asf}. Furthermore, by analyzing the exclusive production of baryon-antibaryon pairs in electron-positron annihilation, BESIII enables precise measurements of nucleon electromagnetic form factors in the time-like region \cite{Denig:2012by}. These studies are crucial for establishing the hadron spectroscopy, nucleon structure and deepening our understanding of confinement in the non-perturbative regime\cite{Wang:2023poi,WANG:2025fmh,Wang:2024whi,Wang:2024xvq,Gao:2026hjv}.

    Photons and nucleons can produce vector mesons and nucleons under high-speed collisions($\gamma N \to VN$) a reaction known as the photoproduction of vector mesons. Numerous theoretical studies have shown that at low energy scales, elastic scattering between vector mesons and nucleons can typically be used to probe the internal properties of nucleons\cite{refn7,refi7,refi8,refn18,refn19}. Within the framework of the Vector Meson Dominance (VMD) model \cite{Kharzeev:1998bz}, the elastic scattering between vector mesons and nucleons in the low-energy region can be linked to the photoproduction process of vector mesons\cite{refn6,refn9,refn11,refn15,refn20,refn21}. This enables us to investigate the gravitational form factors, internal structure, and trace anomaly contributions of nucleons based on the photoproduction of vector mesons\cite{Kharzeev:1998bz,Hatta:2019lxo,Kharzeev:2021qkd,refn8,refn10,refn13,refn16,refn22,refn23,refi21,refi23,refi24,refi26}.

    Indeed, the photoproduction processes of charmonium states such as $J/\psi$ and $\psi(2S)$ are mainly dominated by scalar gluonic operators\cite{refn10}. Therefore, physical quantities that reflect nucleon properties extracted from these processes are more sensitive to the gluon distribution inside the nucleon. For this reason, charmonium photoproduction is considered an important scattering process for studying the internal structure of nucleons. In recent years, facilities like Jefferson Laboratory in the United States have measured cross sections for a series of charmonium photoproduction reactions\cite{refn26,refn27,refn28,refn29,refn30,refn31,refn32,refn33,refn34,refn35}. These measurements provide rare opportunities to explore nucleon internal properties and have promoted a series of theoretical studies\cite{007:2026dow,CLAS:2026lls,Sibirtsev:2004ca,refn37,refn41,refn42,refi20}. Subsequently, future experiments such as the Electron-Ion Collider at Brookhaven National Laboratory (EIC-US) and the planned Electron-Ion Collider for China (EicC) will also offer important experimental data\cite{Accardi:2012qut,Anderle:2021wcy,eiczhong}. However, the current precision and quantity of experimental measurements are still far from sufficient, which limits our ability to extract information about nucleon internal structure from charmonium photoproduction. Therefore, employing various physical models to describe and calculate charmonium photoproduction processes constitutes a vital area of research, offering significant potential to compensate for the lack of high-precision experimental data. Conventionally, vector meson photoproduction is described by Regge theory through the exchange of a soft Pomeron trajectory\cite{Martynov:2002ez}. However, the large mass of charmonia introduces a hard scale that necessitates a perturbative QCD (pQCD) framework. In this regime, the vacuum exchange is resolved into two-gluons in a color-singlet state \cite{refi19,refi20,refi27,refi28,refi29,refi30}. This two-gluon exchange model directly links the scattering amplitude to the generalized gluon distribution of the target, making exclusive quarkonium production a sensitive probe of nucleon structure.

     Experimentally, ultraperipheral collisions (UPCs) at the Large Hadron Collider (LHC) provide an ideal laboratory environment for studying these processes. In UPCs, relativistic ions interact through strong electromagnetic fields, treated as quasi-real photon fluxes, while hadronic interactions are suppressed due to large impact parameters. This configuration enables the study of coherent photon-nucleus interactions, thereby offering unique insights into the behavior of two gluon exchange mechanisms within dense nuclear media.
However, describing these collisions remains challenging. Standard simulations such as STARlight typically rely on the impulse approximation or on phenomenological scaling assumptions \cite{refn43,refn44,refn45,refn46}. As a result, they may overestimate coherent charmonium production in heavy nuclei, especially at midrapidity, where the process probes small values of Bjorken-$x$ and is expected to be sensitive to nuclear modifications of the gluon distribution, including gluon shadowing and possible saturation effects.

In this work, we address these limitations by implementing the elementary two gluon exchange photoproduction cross sections in the STARlight framework and by comparing the resulting UPC predictions with available LHC measurements. Section~\ref{sec2} introduces the theoretical formalism, including the two gluon exchange description of the elementary photoproduction process and the standard UPC factorization in terms of photon fluxes and photon--target cross sections. Section~\ref{sec4} presents the numerical results for $J/\psi$ and $\psi(2S)$ production in Pb-Pb and $p$-Pb collisions, including rapidity and transverse-momentum distributions. The role and limitations of the phenomenological suppression factor are discussed explicitly. Section~\ref{sec5} summarizes the main conclusions.

	\section{Formalism}
    \label{sec2}
    \subsection{Two-gluon exchange model}

\begin{figure}[h]
        \centering
        \includegraphics[width=1\linewidth]{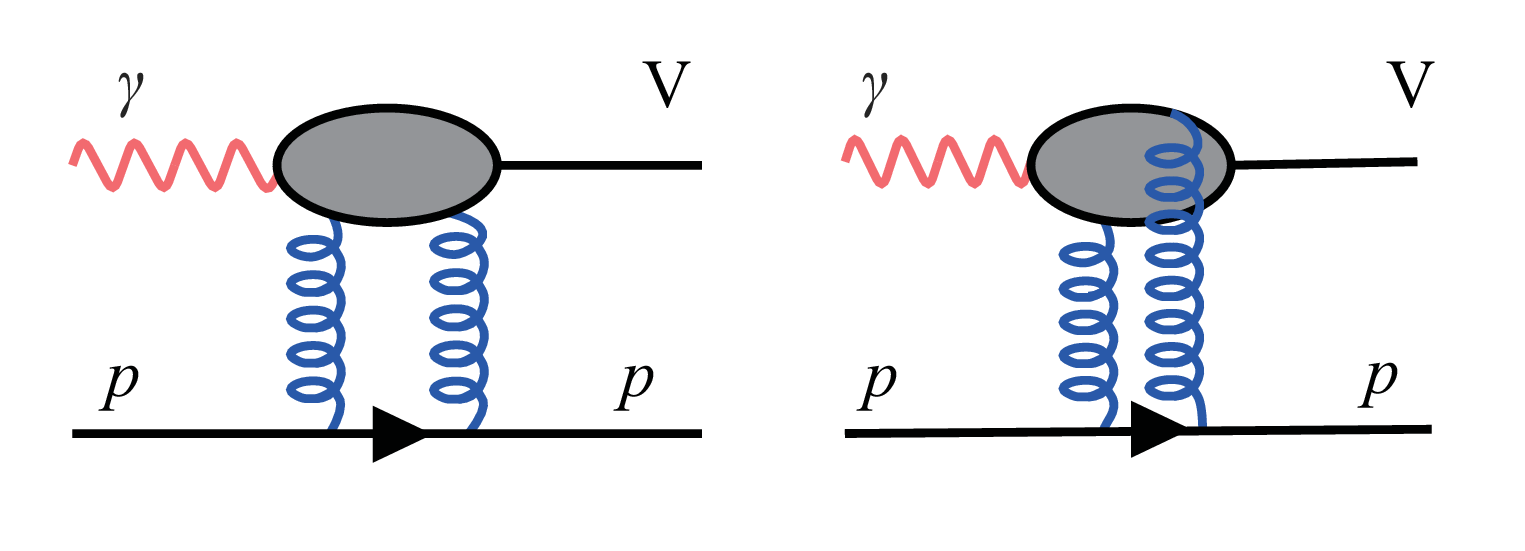}
        \caption{Feynman diagram of the two-gluon exchange model for vector meson photoproduction.}
        \label{fig1}
    \end{figure}

 The two gluon exchange model provides a perturbative-QCD description of exclusive heavy-vector-meson photoproduction. The physical picture is illustrated in Fig.~\ref{fig1}: the photon first fluctuates into a compact $c\bar c$ dipole, which scatters elastically from the proton through the exchange of two-gluons in an overall color-singlet state and then forms the final vector meson $V$ $(V=J/\psi,\psi(2S))$. This color-singlet two-gluon exchange is the lowest-order QCD mechanism compatible with an exclusive diffractive final state.

    Because the charm-quark mass provides a hard scale, the exclusive vector-meson photoproduction amplitude can be factorized, to leading order in the two gluon exchange treatment, into the photon--quarkonium transition, the gluon coupling to the target, and the quarkonium wave-function normalization. With the conventions of Refs.~\cite{refi27,refs2,refi20}, the scattering amplitude can be written as

   \begin{equation}
    \label{eq1}
    \mathscr{T}=\frac{i 2 \sqrt{2} \pi^{2}}{3} m_{q} \alpha_{s} e_{q} f_{V} F_{2 g}(t) \int d l^{2} D_{g}^{2}(l)\left[D_{+}(l)-D_{-}(l)\right] G(l)
    \end{equation}
\noindent where
    \begin{equation}
        \begin{cases}
        D_{-}(l)=(-2m_{q}^{2}-2l^{2})^{-1} \\
		D_{+}(l)=(-2m_{q}^{2})^{-1}.
        \end{cases}
    \end{equation}

\noindent $D_g(l)$ denotes the gluon propagator associated with the exchanged gluon momentum $l$; in the perturbative limit used to connect the amplitude to the integrated gluon density it behaves as $D_g(l)\sim 1/l^2$, while the soft part is effectively absorbed into $G(l)$. $D_{\pm}(l)$ are the propagators of the intermediate quark line in Fig.~\ref{fig1}: $D_{-}(l)$ corresponds to the case in which the two-gluons couple to different quarks in the $c\bar c$ pair, whereas $D_{+}(l)$ corresponds to both gluons coupling to the same quark. The meson decay constant $f_{V}$ is obtained from the leptonic decay width $\Gamma_{e^+e^-}$ via

    \begin{equation}
        f_{V}=\left(\frac{3m_{V}\Gamma_{e^{+}e^{-}}}{8\pi\alpha^{2}e_{q}^{2}}\right)^{1/2}.
    \end{equation}

\noindent The factor $F_{2g}(t)$ encodes the
    $t$-dependence of the two-gluon coupling to the proton and is written as \cite{refi20}

    \begin{equation}
        F_{2g}(t)=\frac{4m_{p}^{2}-2.8t}{4m_{p}^{2}-t}\frac{1}{(1-t/t_{0})^{2}}.
    \end{equation}
\noindent{in which $t_0=0.71$ GeV$^2$ and $m_p$ is the proton mass. $G(l)$ describes the probability for the dipole to absorb a gluon of momentum $l$. It is related to the gluon distribution function $xg(x,Q^{2})$ through}

    \begin{equation}
        xg(x,Q^{2})=\int dl^{2}\frac{G(l)}{l^{2}}, \quad x = \frac{m_{V}^{2}}{W^{2}}
    \end{equation}

    \begin{figure}[htbp]
    \flushleft
    \includegraphics[width=0.95\linewidth]{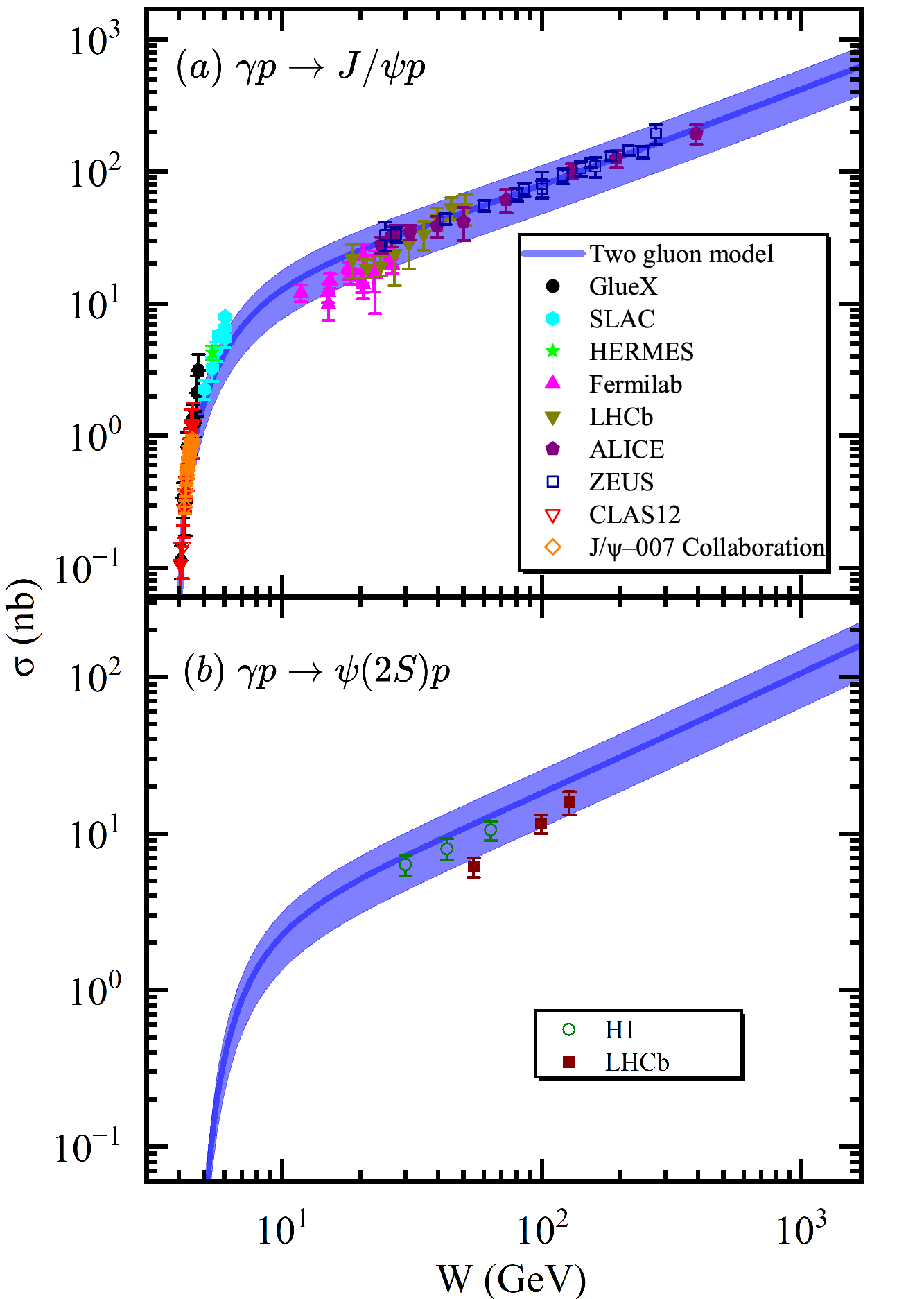}
    \caption{Total cross sections for near-threshold photoproduction of (a) $J/\psi$ and (b) $\psi(2S)$ as functions of the center-of-mass energy $W$. The blue solid curves show predictions from the two-gluon exchange model. Experimental data are from GlueX~\cite{refn26} (black circles), SLAC~\cite{refn29} (blue hexagons), HERMES~\cite{refn34} (green stars), Fermilab~\cite{refn28} (magenta triangles), LHCb~\cite{refn32,refn35} (inverted triangles), ALICE~\cite{refn31,refn33} (purple pentagons), ZEUS~\cite{refn27} (hollow squares), CLAS12~\cite{CLAS:2026lls} (red hollow inverted triangles), $J/\psi$-007 Collaboration~\cite{007:2026dow} (orange hollow rhombus),
    and H1~\cite{refn30,refn34} (hollow circles).}

\label{fig:total_cross_sections}
\end{figure}

\noindent{where $W$ is the center-of-mass energy and $m_{V}$ is the mass of the vector meson. Using this relation, the amplitude of the two-gluon exchange can be recast as}
    \begin{equation}
    \begin{split}
    \mathscr{T} &= \frac{i \sqrt{2} \pi^{2}}{3} m_{q} \alpha_{s} e_{q} f_{V} F_{2 g}(t) \\
    & \left[  \frac{xg(x,Q_{0}^{2})}{m_{q}^{4}}
    \int_{Q_{0}^{2}}^{+\infty}\frac{dl^{2}}{m_{q}^{2}(m_{q}^{2}+l^{2})}\frac{\partial xg(x,l^{2})}{\partial l^{2}}\right].
    \end{split}
    \end{equation}

   \noindent{$Q_0$ denotes the separation scale between the soft input and the perturbatively evolved gluon distribution; in the numerical implementation we take $Q_0\simeq m_V$ following Refs.~\cite{Brodsky:1994kf,refs2,Ryskin:1992ui,Sibirtsev:2003uw}. Using the standard nonrelativistic quarkonium approximation, the forward limit, and the parametrized integrated gluon distribution, the two gluon exchange amplitude leads to the following commonly used expression for the differential photoproduction cross section\cite{refn37,Sibirtsev:2004ca,Sibirtsev:2003uw}: }

    \begin{equation}
        d\sigma/dt =\frac{\pi^{3} \Gamma_{e^{+}e^{-} }{\alpha_{s}} }{6\alpha_{em}m_{q}^{5}} [xg(x,m_{V}^{2})]^{2}e^{bt}.
        \label{eq7}
    \end{equation}

\noindent{Here the gluon distribution is parametrized as $xg(x,m_{V}^{2})=A_{0}x^{A_{1}}(1-x)^{A_{2}}$, with $A_{0}$, $A_{1}$, and $A_{2}$ determined from a fit to photoproduction data~\cite{refs4}. The fitted parameters and the corresponding $\chi^2/\mathrm{d.o.f.}$ are listed in Table~\ref{taba0a1a2}. The exponential slope is taken as $b(W)=b_{0}+0.46\ln(W/W_{\rm thr})$, where $b_{0}=1.67\pm0.38~\mathrm{GeV}^{-2}$ is extracted from near-threshold data~\cite{Hentschinski:2020yfm}. }

\begin{table}[htbp]
\centering
\renewcommand{\arraystretch}{1.5}
\caption{The free parameters $A_0$, $A_1$, and $A_2$ in the gluon distribution function are determined by fitting experimental data and calculating the $\chi^2/\mathrm{d.o.f.}$ value.}
\begin{tabular}{cccc}
\noalign{\hrule height 1.5pt}
$A_0$ & $A_1$ & $A_2$ & $\chi^2/\mathrm{d.o.f}$ \\
\hline

$0.247\pm 0.006$ &$-0.208 \pm 0.004$ & $1.161\pm 0.037$ &$1.949$\\

\hline
\noalign{\hrule height 1.5pt}
\end{tabular}
\label{taba0a1a2}
\end{table}

 The total cross section is obtained by integrating the differential cross section over the kinematically allowed momentum-transfer range from $t_{\min}$ to  $t_{\max}$:

\begin{equation}
\sigma = \int_{t_{\min}}^{t_{\max}} \frac{d\sigma}{dt} dt,
\end{equation}
with $t_{\min}$ and $t_{\max}$ given by
\begin{equation}
t_{\max(\min)} = \frac{m_{V}^{4}}{4W^{2}} - (p_{\gamma} \mp p_{V})^{2}
\end{equation}
while the center-of-mass energies and momenta of the photon and the vector meson are
    \begin{equation}
        \begin{aligned}
           &E_{\gamma}=\frac{W^{2}-m_{N}^{2}}{2W}, \quad E_{V}=\frac{W^{2}+m_{V}^{2}-m_{N}^{2}}{2W} \\
           &p_{\gamma}=E_{\gamma}, \quad p_{V}=\sqrt{E_{V}^{2}-m_{V}^{2}}.
        \end{aligned}
    \end{equation}

 Within this framework, the total photoproduction cross sections for $J/\psi$ and $\psi(2S)$ were computed in our previous work~\cite{refi26}. The theoretical predictions are compared with experimental data in Fig.~\ref{fig:total_cross_sections} (a) for $J/\psi$ and Fig.~\ref{fig:total_cross_sections} (b)  for $\psi(2S)$. The results demonstrate that the two-gluon exchange model gives a good description of the $J/\psi$ data over a wide energy range and can be naturally extended to the production of $\psi(2S)$, confirming the suitability and self-consistency of the model for studying charmonium photoproduction.
            \begin{figure}[htbp]
        \centering
        \includegraphics[width=1\linewidth]{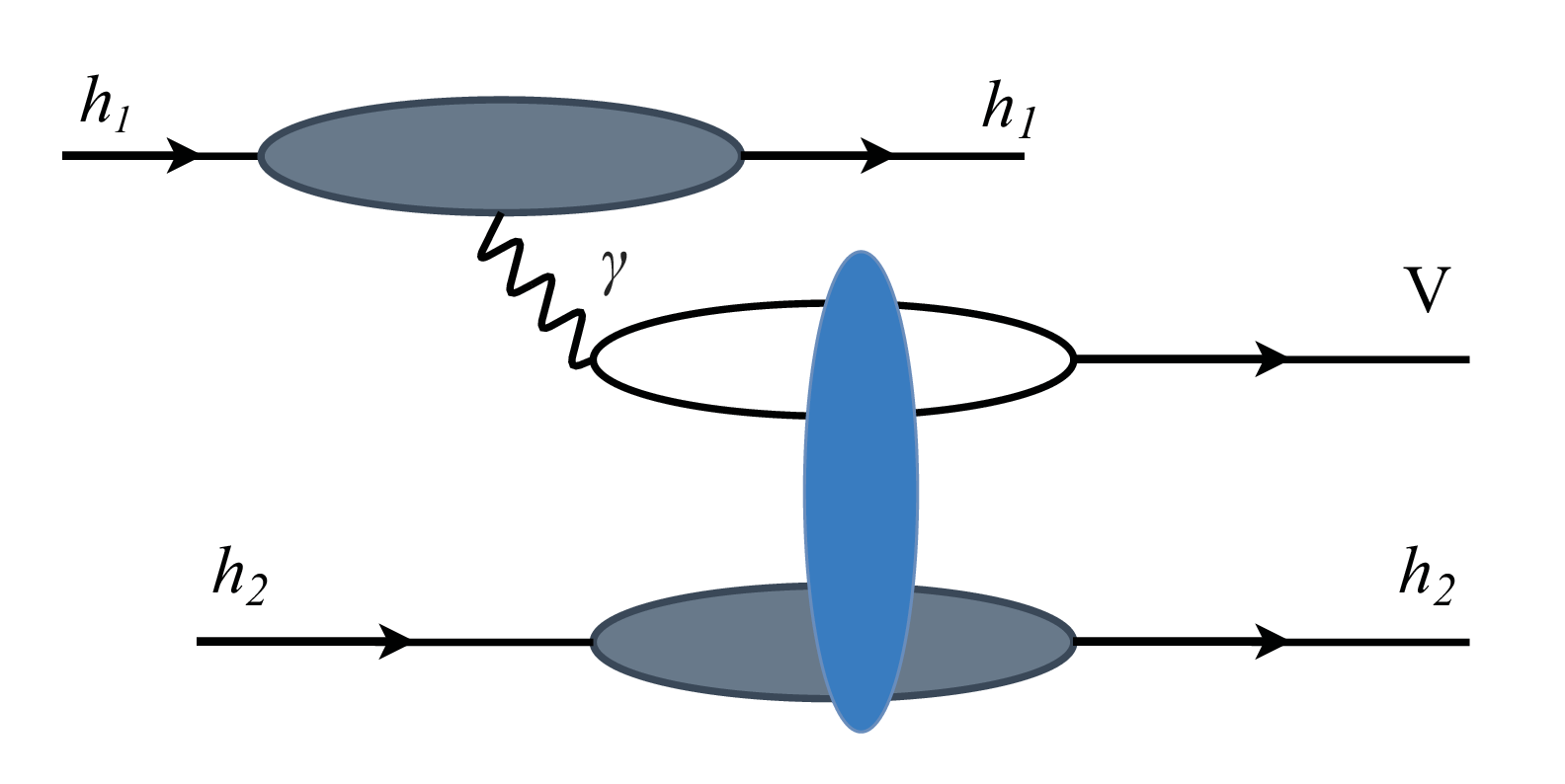}
        \caption{Exclusive vector-charmonium photoproduction in UPCs. The blue blob denotes the coherent photon--nucleus interaction, given by the convolution of the elementary two gluon exchange amplitude with the nuclear form factor.}
        \label{figupc}
    \end{figure}

    \begin{figure*}[htbp]
    \centering
    \includegraphics[width=1.0\linewidth]{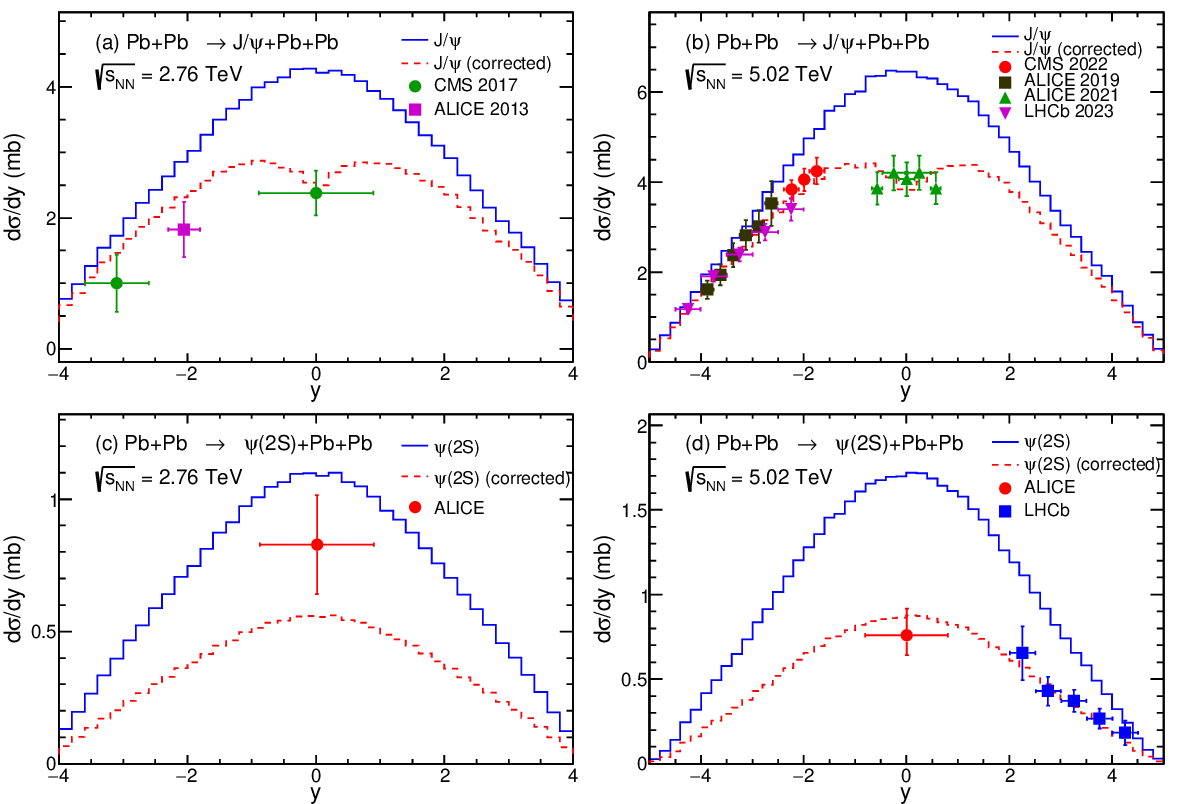}
    \caption{Rapidity distributions for coherent charmonium production in Pb-Pb UPCs: (a) $J/\psi$ at $\sqrt{s_{\rm NN}}=2.76$ TeV, (b) $J/\psi$ at $\sqrt{s_{\rm NN}}=5.02$ TeV, (c) $\psi(2S)$ at $\sqrt{s_{\rm NN}}=2.76$ TeV, and (d) $\psi(2S)$ at $\sqrt{s_{\rm NN}}=5.02$ TeV. The dashed curves show the unmodified STARlight baseline using the two gluon exchange input, while the solid curves include the phenomenological rapidity-dependent suppression factor $\mathcal{F}(y)$. The experimental data are from the ALICE \cite{refu2,refpbpb1,refpbpb2,refpbpb3,refpbpb4,refpbpb5}, CMS \cite{CMS:2023snh,CMS:2016itn}, and LHCb \cite{refpbpb6,refpbpb7} collaborations.}
    \label{fig:placeholder1}
\end{figure*}

	\subsection{The production of vector charmonium at UPCs}
	\label{sec3}

    In hadron--hadron collisions with large impact parameters, the strong interaction between the colliding hadrons is strongly suppressed, while the electromagnetic interaction becomes dominant. Such processes are referred to as ultraperipheral collisions (UPCs) \cite{refn43,refi29}, and the Feynman diagram of this process is shown in Fig.~\ref{figupc}.  In UPCs, hadrons interact via photon-induced reactions, including photon--photon and photon--hadron interactions. In this work, we focus on exclusive vector charmonium photoproduction via photon--hadron interactions.

\begin{figure*}[htbp]
    \centering
    \includegraphics[width=1\linewidth]{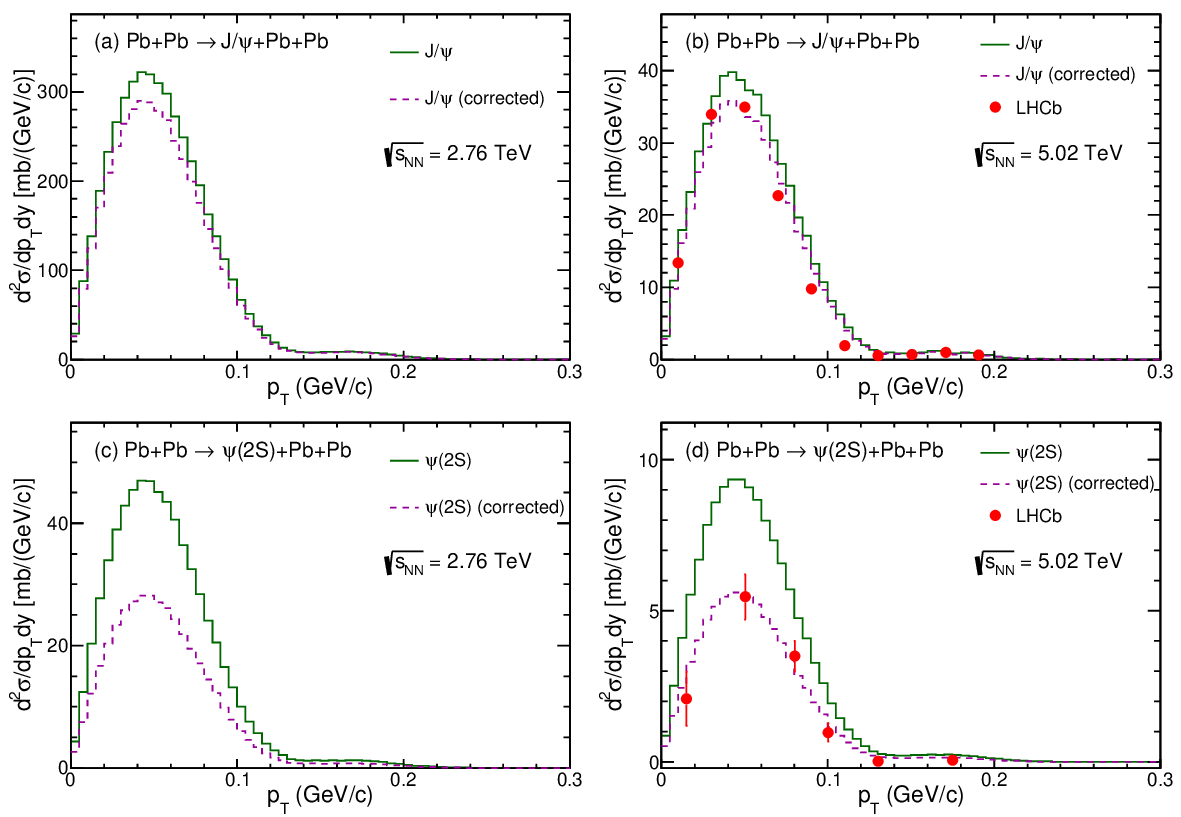}
    \caption{Transverse-momentum distributions for coherent charmonium production in Pb-Pb UPCs: (a) $J/\psi$ at $\sqrt{s_{\rm NN}}=2.76$ TeV, (b) $J/\psi$ at $\sqrt{s_{\rm NN}}=5.02$ TeV, (c) $\psi(2S)$ at $\sqrt{s_{\rm NN}}=2.76$ TeV, and (d) $\psi(2S)$ at $\sqrt{s_{\rm NN}}=5.02$ TeV. The dashed curves include a global normalization factor applied to the STARlight $p_T$ spectrum, approximately 0.90 for $J/\psi$ and 0.60 for $\psi(2S)$. The experimental data points are compared with LHCb measurements where available \cite{refpbpb6}.}
    \label{fig:placeholder2}
\end{figure*}

For relativistic ions, the electromagnetic field generated by the moving charge can be treated as a flux of quasi-real photons within the Weizs\"acker--Williams equivalent photon approximation \cite{refi8}. Consequently, the cross section for vector charmonium production in UPCs can be factorized into the photon flux and the photon--target interaction cross section. Both colliding ions can serve as photon emitters, and the total cross section is obtained by summing the contributions from the two photon sources, which
can be written as\cite{refu3}:

 \begin{equation}
    \begin{split}
        \sigma(h_{1}h_{2}\to Vh_{1}h_{2},\sqrt{s})=
        \int dk\frac{dN}{dk}\big|_{h_{1}} \cdot\sigma_{\gamma h_{2}\to Vh_{2}}(W_{\gamma h_2}) \\
        + \int dk\frac{dN}{dk}\big|_{h_{2}} \cdot\sigma_{\gamma h_{1}\to Vh_{1}}(W_{\gamma h_1}),
    \end{split}
    \end{equation}
    \begin{figure*}[htbp]
    \centering
    \includegraphics[width=1\linewidth]{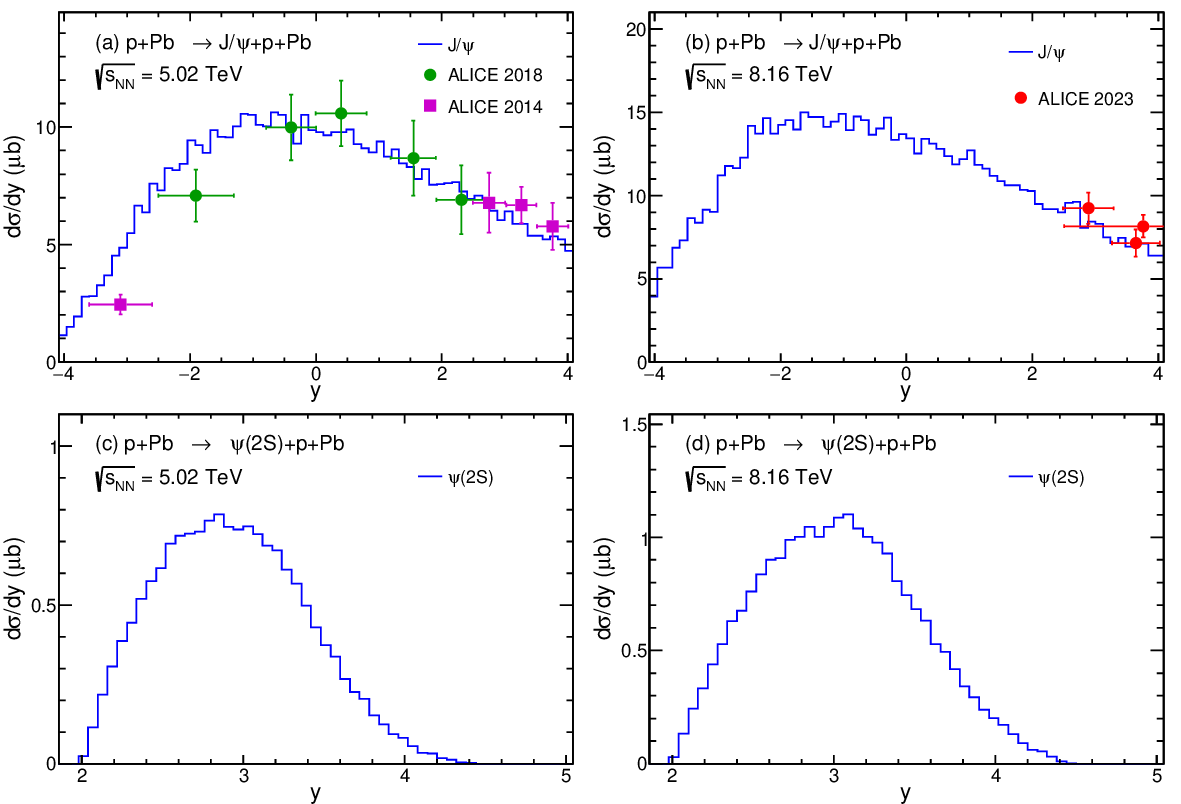}
    \caption{The calculations are presented for (a) $J/\psi$ at $\sqrt{s_{NN}} = 5.02$ TeV, (b) $J/\psi$ at $\sqrt{s_{NN}} = 8.16$ TeV, (c) $\psi(2S)$ at $\sqrt{s_{NN}} = 5.02$ TeV, and (d) $\psi(2S)$ at $\sqrt{s_{NN}} = 8.16$ TeV. The distributions exhibit significant asymmetry due to the unequal photon fluxes from the proton and the lead nucleus. The solid lines represent the model predictions, which are compared with experimental data from ALICE \cite{refn31,refn33}. }
    \label{fig:placeholder3}
\end{figure*}
where $k$ is the momentum of the radiated photon from the hadron and $ \frac{dN(k)}{dk}\big|_{h_{i}} $ represents the photon flux associated with hadron $h_{i}$. The quantity $W$ is the center-of-mass energy of the photon--hadron system, which is related to the photon energy through  $W=\sqrt{2k\sqrt{s}}$ with $\sqrt{s}$ being the center-of-mass energy of the hadron--hadron collision.

For nuclear targets, exclusive photoproduction can occur coherently, in which the photon couples to the entire nucleus and the nucleus remains intact in the final state. The coherence condition constrains the momentum transfer to small values and suppresses contributions from large $|t|$, thereby enhancing the role of nuclear structure effects. Under this assumption, the photon--nucleus cross section can be expressed as

 \begin{equation}
        \sigma(\gamma h\to Vh)=\frac{d\sigma(\gamma h\to Vh)}{dt} \bigg|_{t=0}\int_{t_{\min}}^{\infty}dt\,\left|F(t)\right|^{2}.
    \end{equation}
where $F(t)$ is the elastic form factor of the target. For a nuclear target, $F(t)$ is obtained from the Fourier transform of the nuclear density distribution and carries the dominant small-$|t|$ dependence of coherent production. The elementary photoproduction cross section is evaluated at the forward point, $d\sigma/dt|_{t=0}$, as in the Klein--Nystrand/STARlight implementation. This is justified because the kinematically allowed $|t_{\min}|$ is very small at LHC energies and the elementary exponential dependence, $\exp(bt)$, varies only mildly between $t=0$ and $t=t_{\min}$. By contrast, the form-factor integral must start from $t_{\min}$ in order to respect the physical coherence boundary and to avoid including kinematically inaccessible momentum transfers. In this work, the forward $\gamma p\to Vp$ input is calculated with the two gluon exchange model~\cite{refn53,refu3}.

For phenomenological applications and comparison with experimental measurements, it is often convenient to express the production cross section in terms of the rapidity distribution of the vector charmonium. The rapidity distribution in UPCs can be written as \cite{refu5}

  \begin{equation}
        \begin{split}
            \frac{d\sigma}{dy}=k^{+}\frac{dN}{dk^{+}}\big|_{h2}(k^{+})\sigma^{\gamma h_{1}\to Vh_{1}}(W^{+})+\\
            k^{-}\frac{dN}{dk^{-}}\big|_{h1}(k^{-})\sigma^{\gamma h_{2}\to Vh_{2}}(W^{-})
        \end{split}
    \end{equation}
 where $k^{\pm}=M_{V}e^{\pm y}/2$ and $M_{V}$ denotes the mass of the vector charmonium state. This formulation provides a direct connection between the experimentally accessible rapidity distributions and the underlying production mechanism in UPCs.

\begin{figure*}
    \centering
    \includegraphics[width=1\linewidth]{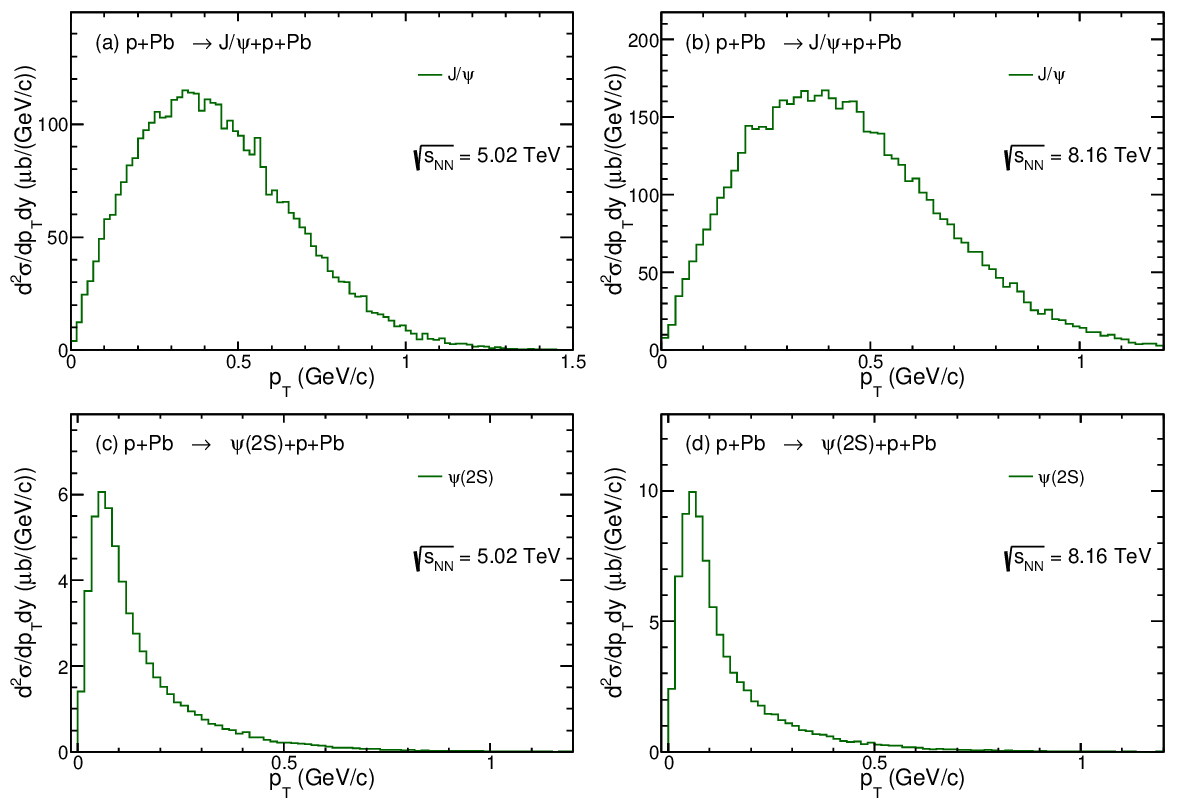}
    \caption{The panels display the theoretical results for (a) $J/\psi$ at $\sqrt{s_{NN}} = 5.02$ TeV, (b) $J/\psi$ at $\sqrt{s_{NN}} = 8.16$ TeV, (c) $\psi(2S)$ at $\sqrt{s_{NN}} = 8.16$ TeV, and (d) $\psi(2S)$ at $\sqrt{s_{NN}} = 5.02$ TeV. The solid curves represent the STARlight calculations using the two gluon exchange elementary input. The low-$p_T$ structure is controlled by the photon transverse momentum and by the elastic form factor of the target; when the Pb nucleus is the target, the nuclear form factor gives the dominant coherent diffractive scale. These results serve as theoretical baselines for future experimental measurements at the LHC.}
    \label{fig:placeholder4}
\end{figure*}

\begin{figure}[htbp]
    \centering
\includegraphics[width=0.95\linewidth]{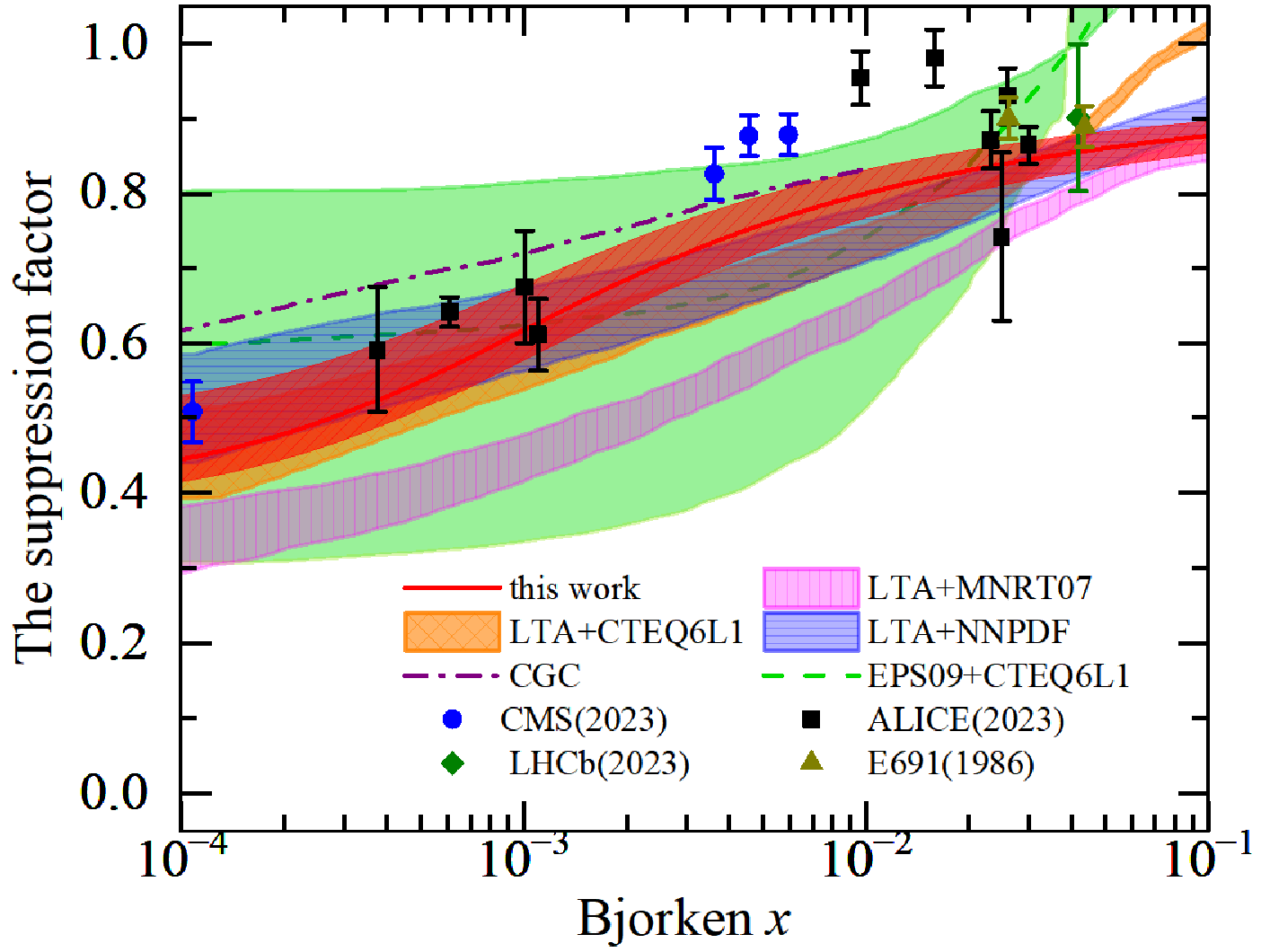}
    \caption{Comparison of the $x$-dependent suppression factor extracted in this work with leading-twist nuclear shadowing calculations, nPDF-based estimates, CGC/saturation-based results, and available experimental extractions. The comparison includes LTA calculations with different proton PDFs \cite{Martin:2007sb,Tung:2002vr,Martin:2006qz,Ball:2010de,Martin:2009iq,Guzey:2024spb}, saturation-based calculations \cite{Mantysaari:2023xcu,Penttala:2025gse}, and experimental information from ALICE \cite{Guzey:2018tlk,refpbpb6}, CMS \cite{CMS:2023snh}, LHCb \cite{refpbpb7}, and E691 \cite{Guzey:2020ntc,FermilabTaggedPhotonSpectrometer:1986xzf}.}
    \label{fig_suppressionfactor}
\end{figure}

\begin{figure}[htbp]
    \centering
\includegraphics[width=0.95\linewidth]{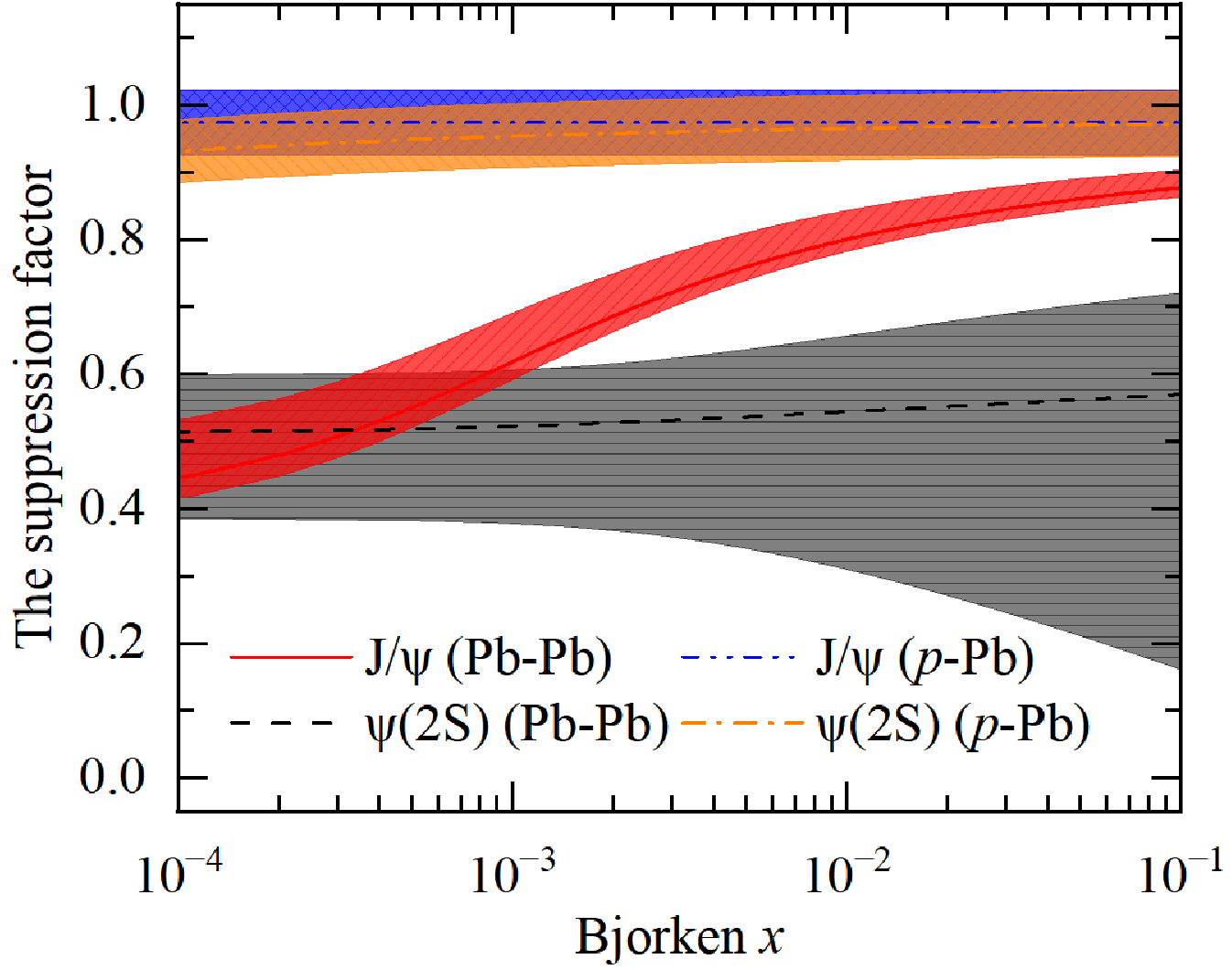}
    \caption{Diagnostic suppression ratios extracted by comparing the model baseline with available data for $J/\psi$ and $\psi(2S)$ production in Pb-Pb and $p$-Pb UPCs. The Pb-Pb ratios indicate a stronger reduction, especially toward small $x$. The $p$-Pb ratios are close to unity and are shown only to illustrate the system dependence; they are not used as an additional phenomenological correction in the $p$-Pb rapidity and $p_T$ predictions. The $\psi(2S)$ curves should be interpreted with larger uncertainties due to limited experimental constraints.}
    \label{fig_suppressionfactor2}
\end{figure}

     \section{Numerical results}
	\label{sec4}	

    In this section, we present the numerical simulations of exclusive vector charmonium photoproduction in ultraperipheral collisions  at LHC energies. The calculations employ the STARlight Monte Carlo generator \cite{refn53}, with photoproduction cross sections $\gamma p \to V p$ from the two-gluon exchange model  serving as the fundamental input, thereby linking the perturbative QCD description of the elementary process to the nuclear photon flux and collision geometry. A total of one million exclusive events were generated for both Pb-Pb and $p$-Pb systems to  ensure simulation accuracy and eliminate statistical fluctuations.
   \subsection{Charmonium Production in Pb-Pb Collisions}
    Fig.~\ref{fig:placeholder1} shows the rapidity differential cross sections $d\sigma/dy$ for coherent $J/\psi$  and $\psi(2S)$  production in Pb-Pb collisions at $\sqrt{s_{NN}} = 2.76$ TeV and $5.02$ TeV.

   The dashed curves show the original STARlight results obtained with the two gluon exchange input. They reproduce the broad rapidity dependence but overestimate the measured cross sections, most visibly around midrapidity ($y\to0$). Similar suppressions relative to the impulse approximation have been observed at RHIC and LHC energies and are commonly associated with small-$x$ nuclear gluon effects \cite{refpbpb8,refpbpb6,refpbpb7}. In the present analysis we do not attempt to derive these effects microscopically. Instead, we introduce a phenomenological rapidity-dependent factor,

    \begin{equation}
    \mathcal{F}(y) = e^{-\frac{1}{a|y|+b}},
    \end{equation}
which parametrizes the reduction required by the data relative to the STARlight baseline. The form of $\mathcal{F}(y)$ is not a microscopic calculation of shadowing or saturation; it is chosen as a compact way to describe a stronger reduction near midrapidity and a weaker reduction at larger $|y|$, as expected qualitatively when the probed Bjorken-$x$ changes with rapidity~\cite{Alvioli:2018krm,Guzey:2013qza,Guzey:2018tlk,Guzey:2024spb}. The free parameters $a$ and $b$ are determined by fitting the rapidity-distribution data; the results are listed in Table~\ref{tabab}. After applying this factor, the solid curves in Fig.~\ref{fig:placeholder1} reproduce both the overall magnitude and the characteristic ``rabbit-ear'' structure of the $J/\psi$ distributions. For $\psi(2S)$, the present data are less precise and have larger uncertainties, so the fitted suppression pattern should be interpreted more cautiously.

\begin{table}[htbp]
\centering
\renewcommand{\arraystretch}{1.5}
\caption{Fitted parameters $a$ and $b$ in the rapidity-dependent suppression factor $\mathcal{F}(y)$, together with the corresponding $\chi^2/\mathrm{d.o.f.}$ values.}
\begin{tabular}{lccc}
\noalign{\hrule height 1.5pt}
Meson & $a$ & $b$ & $\chi^2/\mathrm{d.o.f}$ \\
\hline

\textbf{$J/\psi$} &$1.41 \pm 0.26$ & $1.77\pm 0.21$ &0.665\\
\hline
\textbf{$\psi(2S)$} & $0.06\pm0.16$&$1.48\pm 0.40$&0.704\\
\hline
\noalign{\hrule height 1.5pt}
\end{tabular}
\label{tabab}
\end{table}

Fig.~\ref{fig:placeholder2} displays the transverse-momentum distributions of charmonium in Pb-Pb collisions. The STARlight calculation reproduces the low-$p_T$ diffractive pattern, indicating that the nuclear form factor gives a reasonable description of the coherent scattering geometry and of the transverse spatial distribution of the Pb nucleus. The main discrepancy is in the normalization. We therefore apply a global, rapidity-independent normalization factor to the simulated $p_T$ spectrum, about 0.90 for $J/\psi$ and 0.60 for $\psi(2S)$, obtained by weighting the simulation to the available experimental data. This treatment is separate from Eq.~(14): it adjusts the overall $p_T$ yield while preserving the diffractive shape. Because the available $p_T$ constraints are limited, these constants should be interpreted as empirical normalization factors rather than as a direct integral consequence of $\mathcal{F}(y)$.

\subsection{Charmonium Production in $p$-Pb Collisions}

To investigate the impact of collision system asymmetry on the production mechanism, we also simulated the production of vector charmonium in $p$-Pb collisions.

Fig.~\ref{fig:placeholder3} shows the rapidity distributions for $J/\psi$ and $\psi(2S)$ production in $p$-Pb collisions. In contrast to the symmetric Pb-Pb case, the $p$-Pb distribution is strongly asymmetric because the two photon-induced branches are physically different. When the Pb nucleus acts as the photon emitter, the photon mainly scatters from the proton target ($\gamma p$); when the proton emits the photon, the reaction involves a nuclear target ($\gamma A$). The $\gamma p$ branch is not affected by nuclear gluon shadowing, while the $\gamma A$ branch can be suppressed by nuclear modifications of the gluon distribution. Therefore, the net suppression expected in $p$-Pb collisions is naturally weaker than that in Pb-Pb collisions, where the coherent production process always involves a Pb nuclear target. This explains why the $p$-Pb results are already close to the data without the strong rapidity-dependent correction required in Pb-Pb collisions \cite{Guzey:2018tlk,Guzey:2013qza,Guzey:2024spb}.

The transverse-momentum distributions in Fig.~\ref{fig:placeholder4} lead to the same conclusion. Once the asymmetric photon flux is treated properly, the two gluon exchange input combined with the nuclear form factor gives a reasonable baseline for $p$-Pb UPCs, without invoking a Pb-Pb-like rapidity-dependent suppression.

To further clarify the physical implication of the fitted factor, we convert the rapidity dependence into an approximate Bjorken-$x$ dependence using $x_{\pm}\simeq (M_{V}/\sqrt{s_{\rm NN}})e^{\pm y}$, which is equivalent to the usual relation $x=M_V^2/W_{\gamma N}^2$ for the corresponding photon--nucleon branch. This conversion is useful because Eq.~(\ref{eq7}) shows that the forward charmonium photoproduction cross section scales approximately as $[xg(x,Q^2)]^2$; coherent charmonium production is therefore highly sensitive to the nuclear gluon distribution at small $x$.

Fig.~\ref{fig_suppressionfactor} compares the suppression factor obtained in the present work with several existing descriptions in the region $10^{-4}<x<10^{-1}$. The extracted curve lies in the same general range as the leading-twist nuclear shadowing calculations and recent saturation-based estimates. More importantly, it follows the same qualitative trend: the suppression becomes stronger as $x$ decreases. This is the region where the gluon density in a heavy nucleus grows rapidly and where nonlinear gluon recombination, often described in terms of gluon saturation, can reduce the growth of the coherent production amplitude. Thus, Fig.~\ref{fig_suppressionfactor} does not prove a unique microscopic mechanism, but it shows that the phenomenological reduction required by the data is compatible with established small-$x$ nuclear gluon suppression scenarios \cite{Guzey:2024spb,Mantysaari:2023xcu,Penttala:2025gse}.

\begin{table}[htbp]
\centering
\renewcommand{\arraystretch}{1.5}
\caption{Total cross sections for $J/\psi$ and $\psi(2S)$ production in Pb-Pb and $p$-Pb collisions.}
\begin{tabular}{lccc}
\noalign{\hrule height 1.5pt}
Meson & Collision System & $\sqrt{s_{NN}}$ (TeV) & Cross section \\
\hline

\textbf{$J/\psi$} & Pb-Pb & 2.76  & 22.906~mb \\
                  &       & 5.02  & 38.423~mb \\
\hline
                  & $p$-Pb  & 5.02 & 65.372~$\mu b$ \\
                  &       & 8.16 & 96.428~$\mu b$ \\
\hline

\textbf{$\psi(2S)$} & Pb-Pb & 2.76 & 5.592~mb \\
                    &       & 5.02& 9.709~mb  \\
\hline
                    & $p$-Pb  & 5.02 & 0.963~$\mu b$  \\
                    &       & 8.16  & 1.415~$\mu b$  \\

\noalign{\hrule height 1.5pt}
\end{tabular}

\label{tab:cross_sections}
\end{table}
Fig.~\ref{fig_suppressionfactor2} shows diagnostic suppression ratios for the two collision systems and the two charmonium states. For $J/\psi$ production, the Pb-Pb ratio is visibly below the corresponding $p$-Pb ratio over most of the plotted $x$ range, indicating a stronger reduction in the symmetric heavy-ion system. This behavior is physically reasonable: in Pb-Pb UPCs, both rapidity branches involve coherent production on a Pb target, whereas in $p$-Pb UPCs the photon--proton branch is not subject to the same nuclear gluon modification. The $p$-Pb ratios are therefore shown only as a diagnostic comparison and are not applied as an additional correction in Figs.~\ref{fig:placeholder3} and~\ref{fig:placeholder4}. The $\psi(2S)$ case shows the same qualitative tendency, but its experimental constraints are weaker and the fitted parameters carry larger uncertainties.

Taken together, the rapidity distributions, the $p_T$ spectra, and the two $x$-dependent comparisons indicate that the impulse-approximation baseline overestimates coherent charmonium production in Pb-Pb UPCs. The suppression factor introduced here should be regarded as a phenomenological parametrization of the reduction required by the data. Its magnitude, $x$ dependence, and system dependence are compatible with small-$x$ nuclear gluon suppression, including both leading-twist nuclear shadowing and saturation-based descriptions, but they do not constitute a first-principles proof of a unique microscopic mechanism.

Finally, by integrating the differential distributions obtained from our corrected model, we predict the total cross sections for vector charmonium production in UPCs. These results are summarized in Table~\ref{tab:cross_sections}. It shows that  the cross sections increase significantly with collision energy $\sqrt{s_{NN}}$, driven by the enhanced photon flux and increasing gluon density at higher energies. Additionally, the production rates for $J/\psi$ are consistently an order of magnitude higher than those for $\psi(2S)$, reflecting the differences in their wave-functions and di-lepton branching ratios $\Gamma_{ee}$. These predictions provide a baseline for future experimental measurements at the LHC.

	\section{Summary}

	\label{sec5}

In this work, we have investigated exclusive $J/\psi$ and $\psi(2S)$ photoproduction in ultraperipheral Pb-Pb and $p$-Pb collisions at LHC energies. The elementary $\gamma p\to Vp$ cross sections were calculated in the two gluon exchange framework and then used as input to the STARlight Monte Carlo generator for coherent UPC production.

The impulse-approximation STARlight baseline is found to overestimate the measured coherent charmonium yields in Pb-Pb collisions, especially near midrapidity. To quantify this reduction, we introduced a phenomenological rapidity-dependent suppression factor. With this correction, the calculated Pb-Pb rapidity distributions reproduce the overall magnitude and the characteristic double-peak structure of the $J/\psi$ measurements from ALICE, CMS, and LHCb. The transverse-momentum spectra retain the diffractive pattern expected from coherent scattering off the full nucleus, while the available $p_T$ data require an additional global normalization factor of about 0.90 for $J/\psi$ and 0.60 for $\psi(2S)$. These constants should be viewed as empirical normalizations, not as a microscopic derivation of nuclear suppression.

For $p$-Pb collisions, the calculated rapidity distributions exhibit a clear asymmetry caused by the unequal photon fluxes from the proton and the Pb nucleus and by the different photon--proton and photon--nucleus branches. Unlike the Pb-Pb case, the available $p$-Pb data can be reasonably described without applying a strong Pb-Pb-like rapidity-dependent correction. This is consistent with the fact that a substantial photon--proton contribution is not subject to the same nuclear modification as coherent production on a Pb target.

The fitted Pb-Pb reduction was translated into an approximate Bjorken-$x$ dependence to provide a qualitative comparison with existing descriptions of nuclear gluon suppression. The extracted reduction tends to become stronger toward smaller $x$, which is compatible with the general expectations of leading-twist nuclear shadowing and saturation-based approaches. However, this comparison is phenomenological and does not by itself establish a unique microscopic mechanism or separate shadowing from saturation effects.

Overall, the present study indicates that the two gluon exchange input, combined with phenomenological nuclear-suppression effects, provides a useful baseline for charmonium production in UPCs. The comparison between symmetric Pb-Pb and asymmetric $p$-Pb systems helps constrain the size and system dependence of nuclear effects in coherent vector-meson production. The predicted total cross sections at several LHC energies may serve as reference values for future UPC measurements.

\begin{acknowledgments}
		This work is supported by the National Natural Science Foundation of China under Grant No. 12565018, the Natural Science Foundation of Gansu province under Grant No. 22JR5RA266, and the West Light Foundation of The Chinese Academy of Sciences under Grant No. 21JR7RA201.
	\end{acknowledgments}

\end{document}